# Title:
# Correlated Disorder in van der Waals Heterostructures

## Authors:
Nouamane Laanait[1*], Zhan Zhang[2], Christian Schleputz[2], Ying Liu[5], Michael Wojcik[2], Rachael L. Myers-Ward[3], D. Kurt Gaskill[3], Paul Fenter[4], and Lian Li[5]

## Affiliations:

[1] Center for Nanophase Materials Sciences, Oak Ridge National Laboratory, Oak Ridge, TN 37831

[2] X-ray Science Division, Argonne National Laboratory, Lemont, IL 60639

[3] U.S. Naval Research Laboratory, Washington, DC 20375

[4] Chemical Sciences and Engineering Division, Argonne National Laboratory, Lemont, IL 60639

[5] Department of Physics, University of Wisconsin, Milwaukee, WI 53211

## Corresponding Author*:

Nouamane Laanait

PO Box 2008, MS 6487

Center for Nanophase Materials Sciences, Oak Ridge National Laboratory

Oak Ridge, TN 37831

Phone: 865 574 2988

Email: laanaitn@ornl.gov

**Significance Statement:**
The nature of disorder in physical systems with long-range order is at the forefront of many fields. Of particular interest are those systems whose underlying interactions, either under-constrain or over-constrain the set of possible configurations. We show with X-ray diffraction microscopy that the mixed nature of cohesive interactions (bonded and non-bonded) in van der Waals (vdW) heterostructures, leads to the presence of in-plane correlated structural disorder from one vdW layer to the next, while maintaining nearly perfect crystallographic order along the out of plane direction. These findings indicate that systems with interactions that are of a mixed nature, such as vdW heterostructures, can host unique disorder states.

**Abstract:**
The individual building blocks of van der Waals (vdW) heterostructures host fascinating physical phenomena, ranging from ballistic electron transport in graphene to striking optical properties of $MoSe_2$ sheets. The presence of bonded and non-bonded cohesive interactions in a vdW heterostructure, promotes diversity in their structural arrangements, which in turn profoundly modulate the properties of their individual constituents. Here, we report on the presence of correlated structural disorder coexisting with the nearly perfect crystallographic order along the growth direction of epitaxial vdW heterostructures of $Bi_2Se_3$/graphene/SiC. Using the depth penetration of X-ray diffraction microscopy and scattering, we probed their crystal structure from atomic to mesoscopic length scales, to reveal that their structural diversity is underpinned by spatially correlated disorder states. The presence of the latter induces on a system, widely considered to behave as a collection of nearly independent 2-dimensional units, a pseudo-3-dimensional character, when subjected to epitaxial constraints and ordered substrate interactions. These findings shed new light on the nature of the vast structural landscape of vdW heterostructures and could enable new avenues in modulating their unique properties by correlated disorder.

**Introduction:**
The structural cohesion of van der Waals (vdW) heterostructures relies upon a delicate balance between strong intra-layer bonded interactions and weak inter-layer coupling(1, 2). These systems are intrinsically 2-dimensional (2D) with each vdW unit consisting of one to a few atomic planes that are covalently bonded in-plane, with weak dispersion forces linking one layer to the next (Fig. 1a). The mixed nature of these cohesive interactions facilitates versatile avenues in manipulating the arrangement of vdw heterostructures, ranging from mechanical stacking to van der Waals epitaxy on mismatched materials (3), but also leads to a myriad of structural configurations that profoundly modulate the properties of vdW systems(4), most notably their unique electronic and optical properties (2, 5-9). For instance, in mechanically stacked vdW heterostructures, the in-plane rotational freedom leads to the formation of Moiré patterns(5, 6) that induce electronic superlattices in graphene/hexagonal boron nitride(8), while epitaxial vdW heterostructures, such as $Bi_2Se_3$/graphene/SiC, exhibit Dirac surface states that are locally "gated" by the various structural modes present (10).

In the presence of epitaxial constraints(3) and a two-dimensional confinement of extended defects(11), the mixed nature of cohesive interactions could generate in vdW systems structural configurations that are absent in fully-bonded crystals(12, 13). Van der Waals epitaxy, in particular, is an appealing method to controllably investigate how the rich structural landscape of vdW heterostructures emerges. For instance, removing the constraint of covalent bonding along the growth direction induces crystallographic arrangements in vdW systems that go beyond the conventional tenets of heteroepitaxy(3). Epitaxial vdW heterostructures can also allow one to peer into the very nature of structural disorder in these 2D sytems, where the classical picture of propagating disorder through extended defects, such as dislocations, is significantly altered due to a truncation and confinement of these continuous fields by the van der Waals "gap"(11).

In this work, we investigated the structural arrangements present in model vdW heterostructures of $Bi_2Se_3$ on epitaxial graphene/6H-SiC (0001) (see Methods for samples synthesis). We found that the burgeoning of diverse structural configurations in $Bi_2Se_3$, such as an apparent 12-fold crystallographic symmetry, are underpinned by spatially correlated disorder states. By accessing both local structural correspondences between substrate and vdW layer as well as layer-layer interactions with the depth penetration of hard X-ray diffraction microscopy, we demonstrate that these disorder states originate at the confluence of atomic scale modulations in the strength of vdW layer-layer coupling and ordered nanoscale interactions imposed by the graphene/SiC substrate. These findings hint at a rich picture of structural disorder in epitxial vdW heterostructures that is significantly different than that present in purely covalently bonded systems.

## Results:

**A Modulated van der Waals Gap.** The atomic configuration of $Bi_2Se_3$ and graphene layers along the *c*-axis of SiC was measured by high-resolution surface X-ray diffraction (HRXRD) through a scan of the specular crystal truncation rod of SiC up to a scattering vector magnitude of $|\mathbf{q}| = 5.8$ Å$^{-1}$ (spatial resolution $\Delta \approx \pi/q = 0.5$ Å)(14) (Fig. 1). The sharp $Bi_2Se_3$ (003) Bragg peaks indicate that the quintuple layers (QL) of the film, consisting of a sequence of {Se-Bi-Se-Bi-Se}, are highly ordered along the *c*-axis of the substrate. Structural refinement with the model in Fig. 1a was performed using genetic algorithms(15), where the experimental scattered intensity $I(\mathbf{q})$ is expressed as a convex linear combination of a thickness-dependent surface structure factor $F_{surface}(n, \mathbf{q})$:

$$I(\mathbf{q}) = \sum_n W(n) I(n, \mathbf{q}),$$
$$I(n, \mathbf{q}) = |F_{bulk}(\mathbf{q}) F_{CTR}(\mathbf{q}) + F_{surface}(n, \mathbf{q})|^2 \quad (1.1)$$

where $F_{bulk}$ is the structure factor of 6H-SiC, $F_{CTR}$ is the crystal truncation rod form factor and $F_{surface}(n, \mathbf{q})$ encapsulates the structure factors of surface SiC bilayers, graphene layers, 2 interfacial QLs of Bi$_2$Se$_3$, "bulk" QLs of Bi$_2$Se$_3$, and 2 Bi$_2$Se$_3$ surface QLs, the model is illustrated in Fig. 1A. *W(n)* is a distribution that gives the fractional occupation of Bi$_2$Se$_3$ domains with thickness *n*, and satisfies $1 = \sum_n W(n)$. Determination of the functional shape of *W(n)* and the extrema of *n* are found using differential evolution, a numerical global optimization method (additional details on the structural refinement are presented in the Supplementary Information (SI). The model fits the data with a crystallographic R-factor = 0.067 and predicts atomic displacements for the graphene/SiC interface in agreement with established values(16, 17) (Supplementary Table 1). For Bi$_2$Se$_3$, the data shows the presence of a modulated vdW layer-layer interaction potential across the entire heterostructure: 1) strong graphene-QL1 and QL1-QL2 interactions, manifested through atomic scale variations in the van der Waals gap (Fig. 1b), 2) gradual relaxation in the Bi$_2$Se$_3$ intra-layer separation near the surface, and away from its interface with graphene. Far from these regions, a symmetric compression in the QL-QL distance is found, converging to ~ 2.5 Å in the "bulk" layers, a value that is substantially smaller than QL-QL separation in a Bi$_2$Se$_3$ single crystal (~ 2.8 Å).

**Multi-modal Structural Imaging.** We imaged the influence of SiC surface topography on the local structural modes of Bi$_2$Se$_3$ by the depth penetration of a full-field hard X-ray diffraction microscope (XDM) (see Fig. 1e and Materials and Methods for additional information)(18, 19). Imaging contrast in XDM is **q**-dependent and is consequently tuned as a function of **q** to achieve optimal contrast for the different lattice configurations present in the system. For instance, geometric phase contrast for the buried SIC surface was performed at a scattering vector $\mathbf{q} = \frac{2\pi}{c_{SiC}}(0,0,6-\delta)$, where $c_{SiC}$ = 15.118 Å and $\delta \approx 10^{-1}$. Analysis of the imaging contrast by Fourier optics methods (Supplementary information, Supplementary Fig. 1) was used to determine step heights.

The depth penetration of 10 keV hard X-rays enables XDM imaging of the buried SiC topography, revealing terraces(20), prominent jagged step edges, and some step bunching (Fig. 2a and Supplementary Fig.1 for bare graphene/SiC surface). Analysis of imaging contrast indicates that steps with a single 6H-SiC unit cell height populate the surface topography(21).

The real space configurations of Bi$_2$Se$_3$ are imaged at the 006 Bragg diffraction, with sensitivity to the film structure across the entire thickness of the vdW heterostructure, producing a 2-dimensional spatial distribution of the number of Bi$_2$Se$_3$ QLs, $n_{QL}(x,y)$, where (x, y) indicates a lateral position on the sample, with the surface normal oriented along SiC [001] (Fig. 2b)(21). Analysis of the imaging contrast was performed by modeling the diffracted X-ray signal by the *N*-slit interference model (see SI).

The multimodal data in Fig.2 a,b were acquired at identical *(x,y)* on the sample and reveal that local atomic disorder in Bi$_2$Se$_3$ is induced by the nanoscale boundary conditions imposed by the 6H-SiC surface. We find direct support of a growth mode that proceeds by 2-dimensional nucleation, with a front that laterally expands outwards and isotropically on the SiC terrace (outlined region in Fig 2a,b). When the 2-dimensional growth front encounters a SiC step (height ~ 1.5 QL), it climbs over this barrier but at the expense of tilting in the atomic planes of subsequent layers (Fig. 2c).

Structural disorder in the $Bi_2Se_3$ film, such as lattice tilts (Fig. 2C), is accessible by spatially resolving the elastic diffuse scattering around a $Bi_2Se_3$ Bragg peak. We found that lattice tilts are mostly prevalent at substrate step locations, while on terraces the (001) planes of $Bi_2Se_3$ are in perfect alignment with the (001) planes of single crystal 6H-SiC (Fig. 2d). The direction of lattice tilts fully rotates around SiC [001] with a magnitude in the range ~ 0.03-0.07° (Supplementary Fig. 2). These tilts occur with little cost in energy for the weakly interacting surface layers (Fig. 1b) and in combination with the 2d growth front winding around a substrate step (22), lead to the emergence of pyramidal structures at the surface of $Bi_2Se_3$ (Fig. 2d, Supplementary Fig. 3).

The stronger QL-QL interactions, away from the interface with graphene, enable the coherent propagation of this localized disorder (Fig 1b), as evidenced by XDM imaging (Fig. 2D), inducing an extended defect analogous to the strain field of a screw dislocation in covalently bonded crystals, that promotes the commonly observed $Bi_2Se_3$ spiral growth[23].

We assumed, throughout, that graphene displays no spatial structural variations, justified by characterizations of bare epitaxial graphene/SiC samples (see Materials and Methods). We are unable, however, to directly verify the local influence of structural variations in graphene, if any, on the disorder modes of $Bi_2Se_3$, given that XDM contrast of buried graphene carries significant "cross-talk" from $Bi_2Se_3$, due to overlap in scattering from the $Bi_2Se_3$ surface and the graphene layers (Supplementary Fig. 4).

**Local Correlation Analysis of In-Plane Disorder.**
The in-plane orientations of $Bi_2Se_3$ were measured by a 3-dimensional reciprocal space (RSV) volume of Bragg reflections from the {105} planes (Fig. 3a). The crystal structure of $Bi_2Se_3$ ($R\bar{3}m$ space group) should exhibit 3-fold symmetry (reflections from {105} planes), but lamellar twinning leads to an extra set of peaks (105)' (23). Instead, here we find that $Bi_2Se_3$ films (5 and 18 QLs) exhibit yet an additional set of peaks, producing the overall appearance of 12-fold rotational symmetry (Fig. 3a-c). All of these reflections are interconnected by a non-zero annulus of intensity, indicating the presence of a small number of $Bi_2Se_3$ QLs with an in-plane lattice vector that rotates continuously (over the range $[0,2\pi]$) in the *ab*-plane of 6H-SiC, but with identical lattice vector magnitude and sharing the same *c*-axis.

A 1-dimensional projection of the RSV, overlapped with an azimuthal scan of SiC 012 peaks, shows that $Bi_2Se_3$ 105 coincides with SiC 102 (Fig. 3B, Supplementary Fig. 5). Consequently, QLs of $Bi_2Se_3$ with (105) reflections, hereafter referred to as QL-R0, form an epitaxial relation with graphene with minimal (~3%) lattice mismatch (Fig. 3d). The other $Bi_2Se_3$ in-plane orientation, QL-R30 (105R30 reflection), has a [100] vector that is rotated by 30° around 6H-SiC [001].

The QL-R0 configuration is, however, only a minor (volume) fraction of the $Bi_2Se_3$ film, with a domain occupation relation given by $n_{QL-R30} \geq 5 n_{QL-R0}$, as deduced from the peak intensities of Fig. 3c. Clearly, the orientation distribution of interfacial QLs obeying the above occupation relation is at odds with the conventional rules of heteroepitaxy.

We further investigated the nature of the orientation disorder by XDM imaging the $Bi_2Se_3$ QL orientations present throughout the entire thickness of the film, by acquiring a collection of images, $\{S\phi_i(x,y)\}_i$, of $Bi_2Se_3$ domains whose reciprocal lattice vector $\mathbf{G_{105}}$ differs by $\delta\phi = \phi_i - \phi_0$ from the 105 reflection (Fig. 4a, see Methods). By reconstructing a real space distribution of maximal angular variations, $\delta\phi_{max}$, in the QL-R30 and QL-R0 configurations (Fig. 4b), we found three-fold larger angular deviations in the QL-R30 domains in comparison to QL-R0 domains. This observation suggests that the QL-R0 configuration is restricted to interfacial QLs (Fig. 3d); the close interaction between the latter and graphene (Fig. 1b) considerably constraining $\delta\phi_{max}$.

Additional insight into the orientation disorder present in the QL-R30 configuration is found by defining a statistical correlation measure over $\{S\phi_i(x,y)\}_i$

$$C(\delta\phi,x,y) = \frac{\int_{\mathcal{A}}\left(I_{\phi_0}(x-x',y-y')-\langle I_{\phi_0}\rangle\right)\left(I_{\phi_i}(x-x',y-y')-\langle I_{\phi_i}\rangle\right)dx'dy'}{\left(\int_{\mathcal{A}}\left(I_{\phi_0}(x-x',y-y')-\langle I_{\phi_0}\rangle\right)^2 dx'dy' \int_{\mathcal{A}}\left(I_{\phi_i}(x-x',y-y')-\langle I_{\phi_i}\rangle\right)^2 dx'dy'\right)^{\frac{1}{2}}} \quad (1.2)$$

where $I_{\phi_i}$ is the diffracted intensity at the azimuthal angle $\phi_i$, (x,y) denotes a lateral location on the sample, and the integration is performed over a small neighborhood $A$ around (x,y) (see Methods for a derivation and statistical properties of Eq. (1.1) ). This correlation measure encodes the statistical dependence between the orientation state of a QL in the QL-R30 configuration ($\phi_0$) at spatial position (x,y) and the orientation state ($\phi_i$) of subsequent QLs at the same lateral position (x,y) but at different heights along the c-axis of the vdW heterostructure. $C(\delta\phi,x,y)$ reveals the presence of strong statistical correlation, $C \sim 1$, between the orientation states of QLs that extend to the next-nearest neighbor in the rotation chain (Fig. 4c). Spatial averaging of $C(\delta\phi,x,y)$ produces the orientation correlation function, $\overline{C}(\delta\phi)$ in Fig 4d, which, as a function of $\delta\phi$, falls off slower than an exponential decay, indicating that the observed orientation disorder is non-Markovian but correlated, where each orientation state of a QL(s) retains considerable memory of the previous states (Fig. 4c,d). Moreover, $\overline{C}(\delta\phi)$ is nearly symmetric in $\delta\phi$; establishing a physical equivalence between states with angular deviations $\pm\delta\phi$ with respect to the QL-R30 configuration and suggesting that $Bi_2Se_3$ QLs twist continuously along the growth direction (Fig. 4a).

## Discussion

Similarly to the structural disorder of localized lattice tilting, the in-plane orientation disorder is also correlated along the growth direction of the van der Waals heterostructure, albeit the correlation of the latter is of an angular nature, in contrast to lattice tilting which is spatially localized by nanoscale variations in vdW layer interactions with the substrate. This angular correlation is consistent with the average atomic structure of the vdW heterostructure, which indicates the presence of close interactions between QLs in the "bulk" of the $Bi_2Se_3$ film (Fig. 1b). In both instances, the presence of different types of correlations allows for the coexistence of disorder states with the nearly perfect crystallographic order along the *c*-axis. The azimuthal correlation, in particular, induces a pseudo 3-dimensional character on the entire vdW heterostructure. The effect of this continuous vdW layer twist across the $Bi_2Se_3$ film could influence the electronic structure, as was recently shown for twisted graphene lattices (24).

The presented data does not provide a direct explanation for the presence of two in-plane dominant orientations states of $Bi_2Se_3$ (i.e. QL-R0 and QL-R30) while the conventional rules of epitaxy clearly favor the QL-R0 state. Recent results on other van der Waals heterostructures suggest that the presence of strong interaction between a vdW cluster's edge and the substrate can induce an orientational relation between the two that is no necessarily a minima of the total energy of a system interacting purely through dispersion forces (3). If such "edge" interactions are also relevant in the vdW heterostructures studied here, then they could provide an explanation for the presence of two orientation modes: QL-R0 and QL-R30. In addition to 2D nucleation (evidence for which is given by the multimodal imaging in Fig. 2), edge interactions between $Bi_2Se_3$ and SiC step edges could promote step flow growth with one side of the pyramidal domains aligned to the SiC step edges, thereby introducing a coherent orientation mode that is both different than QL-R0 and energetically stable to propagate to the rest of the system (see Supplementary Fig. 6 for scanning tunneling microscopy data hinting at the presence of this growth mode). The $Bi_2Se_3$ lattice tilts observed near SiC step edges (Fig. 3d) could be an out of plane structural signature of the competition between these two orientations or growth modes. Clearly, additional imaging investigations are needed to elucidate this asymmetry in the rotation states of the system. Nevertheless, our results indicate that depth penetrating, three-dimensional structural imaging is required, to fully capture the character of the complex distribution of disorder states that are present in this model vdW heterostructure and probably many others.

The disorder modes that we demonstrated highlight the rich structural landscape of epitaxial vdW heterostructures and indicate that the principles of vdW epitaxy go beyond those of conventional epitaxy, driven by the inherent flexibility and anisotropy in their cohesive interactions and their interplay with nanoscale boundary conditions imposed by the substrate. Furthermore, they highlight the important role of spatially correlated disorder states in determining the overall structural configurations of these systems. These findings also suggest the feasibility of generating different correlated structural disorder modes in epitaxial vdW heterostructures, by introducing ordered substrate interactions, such as alternating terminations, and by engineering modulated layer-layer interaction profiles through vdW blocks with different types. The presence of correlated disorder in epitaxial van der Waals heterostructures may yet provide unique avenues to modulate their fascinating functionalities (12, 25)$^{26}$ and instantiate novel structural states that are prohibited in fully bonded heterostructures.

**Acknowledgement:**

This work was supported by the Eugene P. Wigner fellowship at Oak Ridge National Lab (NL). PF acknowledges support from the Geosciences Research Program of the Office of Basic Energy Sciences (BES), U.S. Department of Energy (DOE) for XRIM instrument development. Use of the Center for Nanophase Materials Sciences (ORNL), the Advanced Photon Source, and the Center for Nanoscale Materials (Argonne National Lab) was supported by the U.S. DOE Office of Science User Facilities. LL acknowledges support from NSF (DMR-1105839 and DMR-1508560). Work at NRL was supported by the Office of Naval Research.

**Author Contributions:**
NL designed the research and analyzed the data. NL, ZZ, CS, and PF performed x-ray experiments. MW fabricated the x-ray optics. LL and YL grew the $Bi_2Se_3$ samples on graphene/SiC samples provided by RLMW and DKG. NL and LL wrote the manuscript with input from all the authors.

**Materials and Methods:**
**Sample Preparation and Characterization**

The epitaxial graphene (EG) samples were synthesized in a commercial chemical vapor deposition reactor on 8 mm x 8 mm semi-insulating (resistivity > $10^{10}$ ohm cm) (0001) 6H-SiC substrates misoriented by *ca*. 0.41° from [11-20] using Si sublimation process at 1540°C controlled by a mass flow of 10 standard liters per minute of high purity Ar at 100 mbar. Before graphene formation polishing damage on the substrates was removed via etching in high purity $H_2$ at 1540°C, producing step bunching seen in Fig. 2a, which was subsequently changed to Ar *in-situ*. Additional synthesis details can be found in (26). Electronic transport properties on contemporaneous samples of EG formed on similar substrates using this process were n-type, had sheet density of about $5\times10^{12}$ $cm^{-2}$ and mobility of 1000 to 1100 $cm^2$ $V^{-1}$ $s^{-1}$. Epitaxial graphene synthesized in this fashion typically has 1 layer on the SiC terraces and 2 to 3 layers on the narrow step edges as confirmed prior to growth of $Bi_2Se_3$ (Supplementary Fig. 7).

The growth of Bi2Se3 films was carried out on epitaxial graphene/SiC(0001) substrates in an ultrahigh-vacuum (UHV) system (base pressure ~ $1 \times 10^{-10}$ torr) that integrates two molecular beam epitaxy chambers and a low-temperature (5–300 K) scanning tunnelling microscope (STM) (22). For the growth of Bi2Se3, the substrate was held at 275–325 °C and Bi and Se were supplied via separate Knudsen cells at 460°C and 250 °C, respectively. In situ STM imaging was used to monitor the surface morphology and electronic structure and to ensure optimal layer-by-layer spiral growth. The as-grown film exhibited a Dirac point 250 meV below the Fermi level in tunneling spectra, indicating the n-type doping typically seen in MBE-grown materials due to Se vacancies (22). Crystal rocking curve at 00<u>18</u> peak shows a full width at half max ~ 70 arc second, indicating high quality thin-films of $Bi_2Se_3$.

**High-resolution Surface X-ray Diffraction and Reciprocal Space Volume:**
Surface diffraction data was acquired with an X-ray energy of 15.5 keV in sector 33-BM at the Advanced Photon Source (Fig. 1C and Supplementary Fig.1). The specular rod provides information on the occupancy of atomic planes and displacements of the vdW heterostructure relative to 6H-SiC single crystal surface (Fig 1A-B). The interference between the different crystallographic structure factors masks the peaks of the graphene layers that are prominent in measurements on bare epitaxial graphene/6H-SiC systems (Supplementary Fig. 4). The anisotropic spatial distribution of $Bi_2Se_3$ QLs is included in the model to properly account for the truncation rod intensity at all scattering vector magnitudes. The model used in the structural refinement is comprised of 4 surface bilayers of SiC, 2 graphene layers, 2 interfacial QLs of $Bi_2Se_3$, "bulk" QLs of $Bi_2Se_3$, and 2 $Bi_2Se_3$ surface QLs (in Fig. 1A). The model is described by the structure factor, $F(\mathbf{q})$, given by

$$F(n,\mathbf{q}) = F_{6H\text{-}SiC}(\mathbf{q})F_{CTR}(\mathbf{q}) + F_{surface}(n,\mathbf{q}),$$
$$F_{surface}(n,\mathbf{q}) = F_{SiC\text{-}bilayers}(\mathbf{q}) + F_{graphene}(\mathbf{q}) + F_{Bi_2Se_3}(n,\mathbf{q}). \quad (1.$$

where $F_i$ is the crystal structure factor of unit cell $i$ (e.g. bulk SiC, etc …), $F_{CTR}(\mathbf{q})$ is the crystal truncation rod form factor, and $n$ is the number of (bulk) quintuple layers of $Bi_2Se_3$ that are sampled by the x-ray beam during the truncation rod measurement. The anisotropic spatial distribution of $Bi_2Se_3$ QLs must be included in the model to properly account for the truncation rod intensity at all scattering vector magnitudes |**q**|. This is clearly seen from the decay of Kiessig fringes away from the $Bi_2Se_3$ 00$\underline{15}$ diffraction spots. Averaging (incoherent) of the $n$-dependent intensity $I(n) = |F(n,\mathbf{q})|^2$, with $F$ given by (1.1) was performed to produce the total intensity,

$$I(\mathbf{q}) = \sum_n W(n) I(n,\mathbf{q}) \quad , (1.3)$$

where $W(n)$ is a distribution that gives the fractional occupation of $Bi_2Se_3$ domains with thickness $n$, and satisfies $1 = \sum_n W(n)$. Determination of the functional shape of $W(n)$ and the extrema of $n$ (Supplementary Fig. 8) are part of the structural refinement that compares the modeled intensity (1.2) to the experimental intensity (described in the SI). The model fits the data with a crystallographic R-factor = 0.067 using the genetic algorithm of differential evolution. Additional details on the structural refinement and statistical analysis are discussed in the supplementary information.

The full 3-dimensional reciprocal space volume of the diffraction zone of $Bi_2Se_3$ in Fig. 3 was reconstructed by collecting a series of area detector scans at a fixed normal scattering vector in reciprocal space while continuously varying the transverse scattering vector by an azimuthal rotation of the sample.

**X-ray Diffraction Microscopy**
Imaging was performed at the Advanced Photon Source using the XRIM instrument at Sector 33-ID-D with an X-ray energy of 10 keV in Bragg geometry(18). A Fresnel zone plate with 60 nm outermost zone width and 150 $\mu$m diameter was used as an objective lens. A CMOS camera with 6.5 $\mu$m pixel (Andor Neo) coupled to a CsI scintillator and 20X optical magnification was used as detector and positioned 1.4 m from the sample. The imaging setup produces an effective pixel size on the sample of ~ 15 nm.

Imaging contrast in XDM is **q**-dependent. For instance, geometric phase contrast for the buried SIC surface was performed at a scattering vector $\mathbf{q}=\frac{2\pi}{c_{SiC}}(0,0,6-\delta)$, where $c_{SiC}$ = 15.118 Å and $\delta \approx 10^{-1}$. Analysis of the imaging contrast by Fourier optics methods (Supplementary information, Supplementary Fig. 1) is used to determine step heights.

Imaging of spatial variations in the number of $Bi_2Se_3$ QLs, $n_{QL}(x,y)$, was performed at $\mathbf{q}=\frac{2\pi}{c_{Bi_2Se_3}}(0,0,6)$, where $n_{QL}(x,y)$ was obtained from an analysis of the thin-film diffraction intensity (Supplementary information, Supplementary Fig. 9 ). Imaging of lattice tilts of $Bi_2Se_3$ was achieved by offsetting the scattering vector $\mathbf{q}=(\frac{2\pi}{a_{Bi_2Se_3}}\delta_H, \frac{2\pi}{a_{Bi_2Se_3}}\delta_K, \frac{2\pi}{c_{Bi_2Se_3}}.6)$ by an amount $\delta_{H,K}$ (sequentially) in the interval $[-3\times 10^{-2}, 3\times 10^{-2}]$, where c = 28.69(2) Å and a = b = 4.13(4) Å. Due to the projective nature of X-ray surface diffraction in reflection (Bragg) geometry, raw images of SiC and $Bi_2Se_3$ have different aspect ratios. In Fig. 2, the $Bi_2Se_3$ images have all been scaled so that they have the same aspect ratio as SiC images.

XDM images acquired during an azimuthal scan $\{S\phi_i\}_i$ have been registered and transformed to a common angular frame of reference (Supplementary Information). This correction is necessary before extracting physically meaningful properties. The image collection $\{S\phi_i\}_i$ was taken by varying the azimuthal angle with a step size of 0.2°, over a range of 10° centered around $\phi_0$ (i.e. the azimuthal angle at the 105 diffraction peak of QL-R30 or QL-R0, Fig. 3c). Due to the depth penetration of hard X-rays, $\{S\phi_i\}$ probes QL orientations throughout the entire thickness of the film and forms a 2d-projected real space image, $S\phi_i(x,y)$ of $Bi_2Se_3$ domains. The recorded diffraction intensity $I_{\phi i}(x,y)$, at a pixel in an image $S\phi_i$, originates from a $Bi_2Se_3$ QL(s): 1) whose $\mathbf{G}_{105}$ vector coincides with the **q**-vector of the microscope, 2) is located at a lateral position *(x,y)*, 3) whose height could be anywhere within the thickness of the film (along SiC[001]).

**Local Correlation Analysis**
The maximal angular deviation maps in Fig 4.b were computed by choosing a particular *(x,y)* in $\{S\phi_i(x,y)\}_i$, and searching along the $\phi_i$-dimension for the angular value, $\phi_{max}$, where the intensity is maximum. The maximal angular deviation of a $Bi_2Se_3$ domain at *(x,y)* is then given by $\delta\phi_{max} = \phi_{max} - \phi_0$.

We are interested in deriving a statistical measure to quantify the interdependence between QLs located at (x,y) with $\phi_o$ and those QLs with orientation $\phi_i$ at the same (x,y) but at different height position. These two QL orientations have associated intensities $I_{\phi_o}(x,y)$ and $I_{\phi i}(x,y)$ that are sampled (belong) to two distinct XDM images, $S\phi_o$ and $S\phi_i$, respectively. If we consider $S\phi_o$ and $S\phi_i$ to be two random multivariates, a measure that captures statistical dependence between the two is naturally given by statistical signal correlation(27), C. In the case that $C(S\phi_o, S\phi_i) \sim 1$ implies that $S\phi_i$ can, in principle, be inferred given $S\phi_o$. Recall that if the intensity value $I_{\phi i}(x,y)$ can be deduced given $I_{\phi o}(x,y)$, implies that the orientation ($\phi_i$) of QLs at *(x,y)* is not statistically independent from orientation ($\phi_o$) of QLs at *(x,y)* and consequently the orientation disorder is not purely stochastic. One can straightforwardly derive this statistical correlation purely from physical arguments as follows. Non-vanishing intensities $I_{\phi i}$ and $I_{\phi o}$ at $(x,y)$ are necessary to indicate the presence of domains at $(x,y)$ with orientations $\phi_i$ and $\phi_o$, therefore the product the product $I_{\phi i}(x,y)$ $I_{\phi o}(x,y)$ must enter into this statistical measure. However, since the scattered intensity, $I_{\phi i}$, scales quadratically with the number of diffracting QLs, we must subtract the corresponding mean intensity $<I_{\phi i}>$ in the image $S\phi_i$ from this product to not bias this measure to thickness dependent fluctuations. Consequently, we arrive at a measure given by

$$C(\delta\phi,x,y) \propto \int_A (I_{\phi_0}(x-x',y-y') - \langle I_{\phi_0}\rangle)(I_{\phi_i}(x-x',y-y') - \langle I_{\phi_i}\rangle) dx'dy',$$

where we average the intensities over an area $A$ given by a 3 x 3 (pixels) neighborhood that is centered around $(x,y)$ and $\delta\phi = \phi_i - \phi_0$. Averaging is necessary to remove the influence of experimental noise and weak intensities. Furthermore, subtracting the mean intensity is needed to normalize this statistical measure as follows,

$$C(\delta\phi,x,y) = \frac{\int_A (I_{\phi_0}(x-x',y-y') - \langle I_{\phi_0}\rangle)(I_{\phi_i}(x-x',y-y') - \langle I_{\phi_i}\rangle) dx'dy'}{\left(\int_A (I_{\phi_0}(x-x',y-y') - \langle I_{\phi_0}\rangle)^2 dx'dy' \int_A (I_{\phi_i}(x-x',y-y') - \langle I_{\phi_i}\rangle)^2 dx'dy'\right)^{\frac{1}{2}}} \quad (1.3)$$

It is straightforward to see that $C(\delta\phi,x,y) = 1$ when $\phi_i = \phi_0$ (autocorrelation is 1) and otherwise $C(\delta\phi,x,y) \leq 1$. Furthermore, the above measure is easily seen to be formally equivalent to a spatially dependent statistical correlation between two random multivariates (i.e. $S\phi_i$ and $S\phi_o$) (27). When computing this measure, the integrals are converted into discrete sums over pixel positions in the appropriate image, and are referenced with respect to $(x,y)$. In the main text, $S\phi_o$ is kept fixed (at the nominal orientation of the QL-R30 configuration as described earlier), while $S\phi_i$ is varied continuously in increasing (or decreasing) order, with consecutive images $S\phi_i$ obtained through an azimuthal rotation of the sample, and therefore probe the spatial distribution of QLs whose [100] direction is rotated by $\delta\phi = \phi_i - \phi_0$ from the [100] direction of QLs with orientation $\phi_o$. We define the correlation function $\overline{C}(\delta\phi)$ by averaging over $(x,y)$. The associated error bars in Fig. 4D are found by computing (1.2) as a function of discrete values of the area $A$, then taking the standard deviation of this set to indicate fluctuations in the mean that were introduced during the averaging over $(x,y)$. The behavior of this correlation function as a function of angular variation $\delta\phi = \phi_i - \phi_0$, indicates how far in angle does a QL rotate while keeping memory of the angular state of previous state as discussed in the main article.

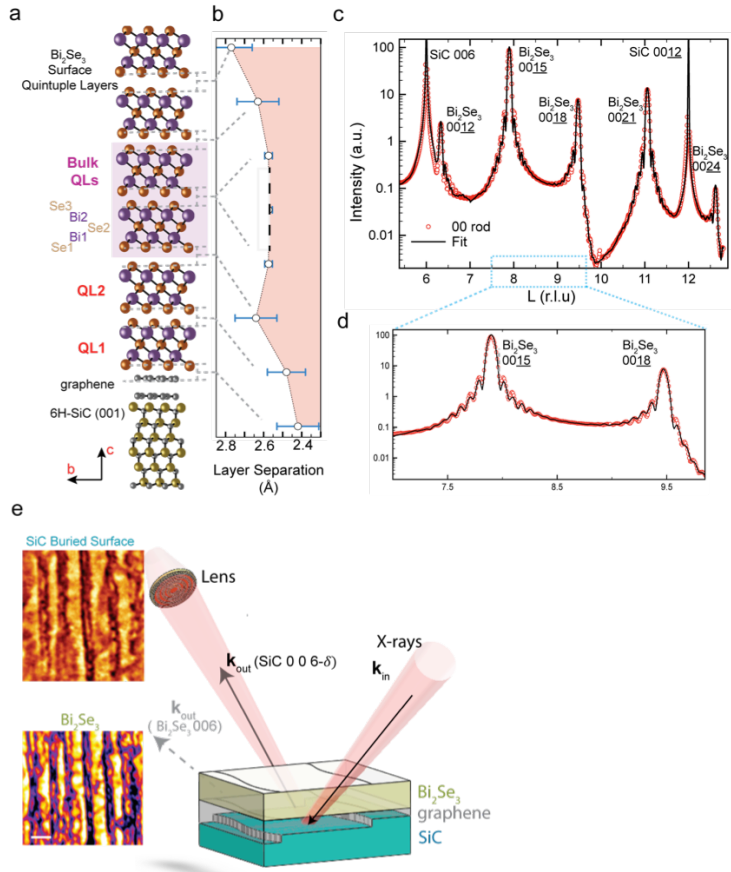

Figure 1. **Multi-Scale imaging of the crystal structure of $Bi_2Se_3$/graphene/6H-SiC (001).** (a) Structural model of the van der Waals heterostructure consisting of a bilayer graphene on SiC and quintuple layers ( 18 QL) of $Bi_2Se_3$. The van der Waals gap variation across the entire height of the heterostructure is indicated in (b) as determined from the structural refinement of the high resolution X-ray diffraction data shown in (c). Error bars in (b) indicate statistical variations in the predictions of the different models used in the structural refinement(21). The specular truncation rod is displayed as a function of the L Miller index, in reciprocal lattice units (r.l.u) of SiC ($|\mathbf{q}| = 2\pi/c_{SiC} L$, $c = 15.113$Å). A close-up view of some of the $Bi_2Se_3$ Bragg peaks in (d) shows both the coherent scattering of the film due to well-ordered layers along [001] (intensity fringes) and the agreement between the structural model and the data (crystallographic R-factor = 0.067). Data error bars are smaller than the symbol size. (e) The depth penetration of hard X-rays (10 keV) facilitates X-ray diffraction microscopy with sub-100 nm lateral resolution of the different lattice configurations present in the system (e.g. buried SiC surface, $Bi_2Se_3$ lattice planes) by tuning the scattering vector $\mathbf{q} = \mathbf{k}_{out} - \mathbf{k}_{in}$ (Methods). Scale bar 1 $\mu$m.

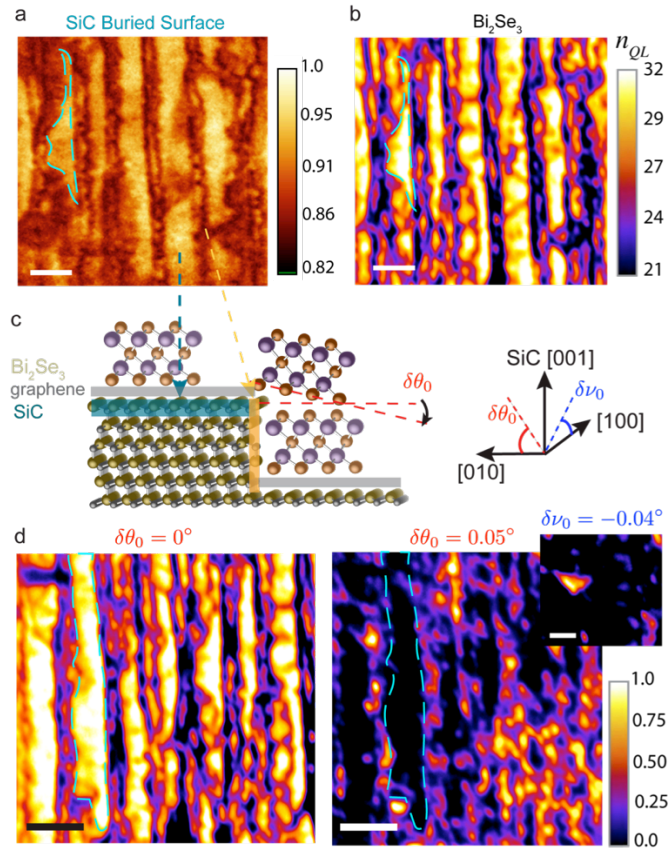

**Figure 2. Multi-modal imaging of the local structural disorder induced on Bi$_2$Se$_3$ layers by interactions with the 6H-SiC surface.** (a) Hard X-ray microscopy image of the buried substrate topography. XDM produces geometric phase contrast for the buried SiC topography (20) with bright contrast for SiC terraces and dark contrast for steps (scale bar 1 μm, colour bar indicates geometric phase contrast). At the same physical location on the sample (dashed outline is a reference), the scattering condition is tuned to Bi$_2$Se$_3$ 006 reflection to image the spatial density of the film's quintuple layers $n_{QL}$ (b), and its dependence on SiC topography (Fig. 1e). (c) Illustration of Bi$_2$Se$_3$ lattice plane tilting and directions ($\delta\theta, \delta\nu$) with respect to SiC crystallographic directions. (d) Configurations of Bi$_2$Se$_3$ domains with lattice tilts are maximal near step edges of SiC and minimal on terraces (scale bar 2 μm, colour bar indicates contrast), acquired by imaging the elastic diffuse scattering around 006 reflection, and suggest a mechanism by which Bi$_2$Se$_3$ achieves spiral growth, in the absence of dislocations, resulting in pyramidal domains at the surface (far right image, scale bar 300 nm).

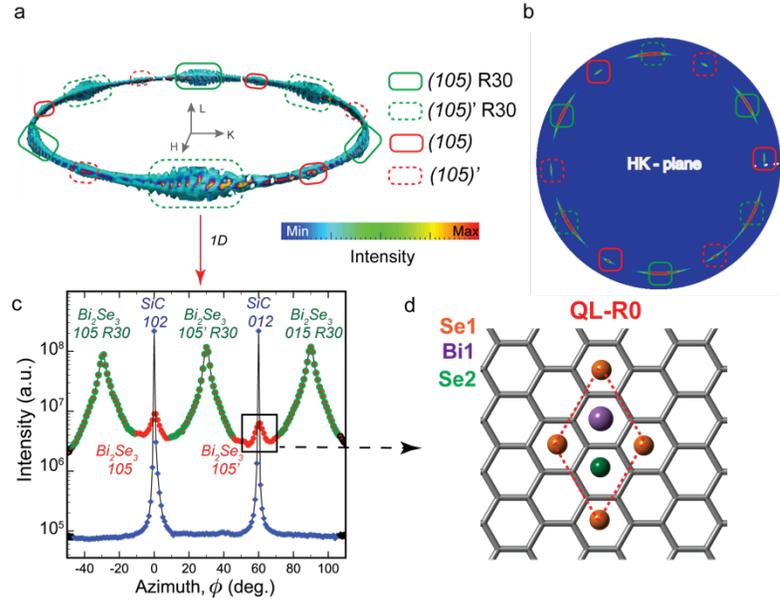

**Figure 3. Two dominant in-plane orientations of $Bi_2Se_3$ on epitaxial graphene/SiC**. The 3-dimensional reconstructed reciprocal space volume from Bragg reflections by the $Bi_2Se_3$ {105} planes (a) indicates the presence of two dominant in-plane structural configurations of $Bi_2Se_3$ in 18 QL $Bi_2Se_3$ and 5 QL $Bi_2Se_3$ (b) with an apparent 12-fold symmetry (c). The QL-R0 orientation have peaks labeled by (105) (and twins with (105)') and QL-R30 with reflections (105) R30 and (105)' R30. QL-R0 have an epitaxial relation given by [100] ∥ SiC [100] and [010] ∥ SiC [010] and form an epitaxial relation with graphene indicated in (c) with minimal lattice mismatch, given that epitaxial graphene on SiC exhibits a $6\sqrt{3}\times6\sqrt{3}R30°$ surface reconstruction[19]. Domains in a QL-R30 configuration have a [100] vector that is rotated by 30° around SiC [001].

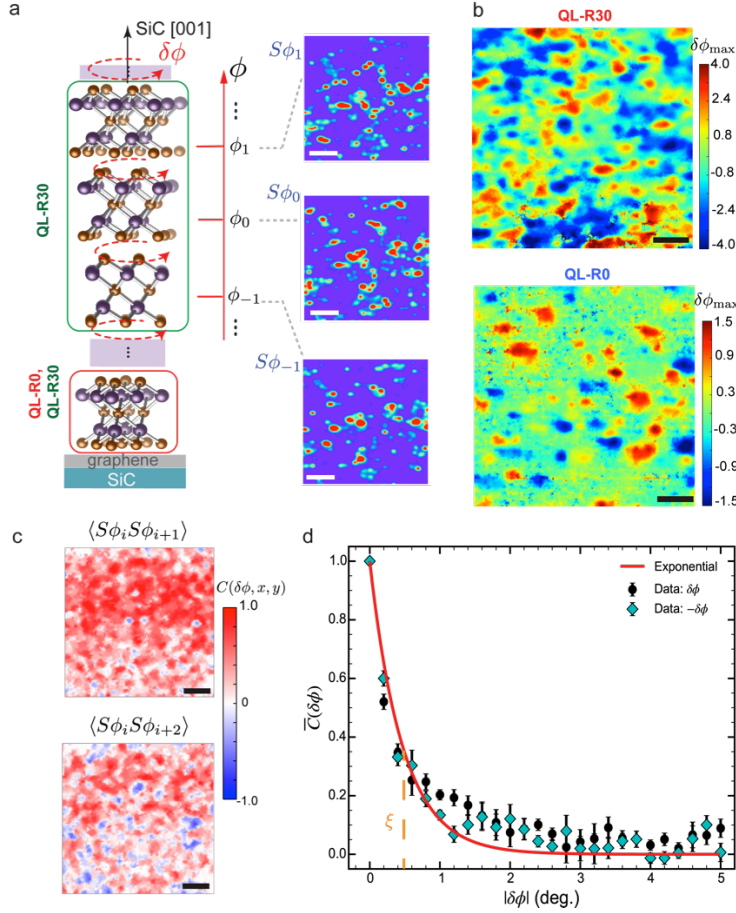

**Figure 4. Local spatial correlations in the orientation disorder of $Bi_2Se_3$.** (a) Illustration of the in-plane arrangement of $Bi_2Se_3$ on epitaxial graphene /6H-SiC(0001), indicating the continuous rotation of $Bi_2Se_3$ QLs about SiC [001] (twist angle, $\delta\phi$) implied by the local correlation analysis in (c and d). The orientation states of QLs, denoted by $\phi_i$, can be considered as a continuous (Markov) chain, where the angular deviation from the QL-R30 (R0) configuration is $\delta\phi = \phi_i - \phi_0$, and $\phi_0$ is the azimuthal angle at the diffraction peaks (Fig. 3c, 105R30 or 105). Representative XDM images, $S\phi_i$, associated with a particular orientation $\phi_i$ are shown in the right of panel (a). (b) Spatial variations of the maximal angular variation present in the QL-R30 (top) and QL-R0 (bottom) configurations, were reconstructed from an XDM collection $\{S\phi_i\}_i$ acquired over an angular range of $10°$. The colour scale indicates the maximal angular rotation $\delta\phi_{max}$ (deg.) of the [100] vector of a QL about SiC [001]. (c) Spatial correlations maps between nearest neighbors (top) orientation states and next-to-nearest neighbors (bottom) states in the angular rotation chain in (a). (d) Spatially averaged orientation correlation function, showing that $Bi_2Se_3$ QLs (belonging to QL-R30 configuration), have an angular state that retains memory of preceding states, over an angular range that exceeds the correlation angle $\xi \approx 0.5°$ in a purely Markovian process (exponential decay). Scale bars are 1.5 $\mu$m.